\documentclass[12pt]{elsart}
\usepackage{amsmath,amsfonts,amssymb,epsfig,color}

\newcommand{\SU}[2]{\ensuremath{\mathrm{SU}^{ #1 }( #2 )}}
\newcommand{\Un}[2]{\ensuremath{\mathrm{U}^{ #1 }( #2 )}}

\newcommand{\Spn}[1]{\ensuremath{\mathrm{Sp}( #1 )}}

\newcommand{\spn}[1]{\ensuremath{\mathfrak{sp}( #1 )}}

\newcommand{\dimN}{\ensuremath{{\mathcal N}}}

\newcommand{\IASs}{isobaric analog $0^+$ states}
\newcommand{\CDB}{``CD-Bonn"}
\newcommand{\CDBt}{CD-Bonn+3terms}
\newcommand{\Gm}{GXPF1}
\newcommand{\HQ}{\ensuremath{H_Q^\perp(2)}}

\newcommand{\Hsp}{\ensuremath{H_{\spn{4}}}}

\newcommand{\HM}{\ensuremath{H_{M}}}

\newcommand{\fpg}{\ensuremath{1f_{5/2}2p_{1/2}2p_{3/2}1g_{9/2}} }

\newcommand{\flevel}{\ensuremath{1f_{7/2}} }
\newcommand{\fFive}{\ensuremath{1f_{5/2}} }
\newcommand{\plevels}{\ensuremath{2p_{1/2}2p_{3/2}} }
\newcommand{\glevel}{\ensuremath{1g_{9/2}} }
\newcommand{\upfp}{upper {\it fp}}

\begin{document}

\centerline{ 
\hskip 4in
\begin{tabular}{r}
UCRL-JRNL-222549 \\
SLAC-PUB-11903
\end{tabular}
}

\begin{frontmatter}

\title{
Global Properties of $fp$-Shell Interactions\\
in Many-nucleon Systems
}
\author{K. D. Sviratcheva$^{a,}$}\footnote{Corresponding author.\\
\textsl{E-mail address:} kristina@baton.phys.lsu.edu; \\
\textsl{Phone:} +225-578-0351;
\textsl{Fax:} +225-578-5855}
\author{, J. P. Draayer$^a$, and J. P. Vary$^b$}
\address{$^a$Department of Physics and Astronomy, Louisiana State University,\\
Baton Rouge, Louisiana 70803, USA \\
$^b$Department of Physics and Astronomy, Iowa State University, \\
Ames, IA 50011, USA \\
Lawrence Livermore National Laboratory, Livermore, California 94551\\
Stanford Linear Accelerator Center,
Stanford University,\\ Stanford, California 94309}
\date{\today}

\maketitle

\begin{abstract}
Spectral distribution theory, which can be used to compare microscopic
interactions over a broad range of nuclei, is applied in an analysis of two
modern effective interactions based on the realistic CD-Bonn potential for
$0\hbar\Omega$ no-core shell model calculations in the $fp$ shell, as well as
in a comparison of these with the realistic shell-model GXPF1 interaction.
In particular, we explore the ability of these interaction to account for the
development of isovector pairing correlations and  collective  rotational
motion in the $fp$ shell. Our findings expose the similarities of these
two-body interactions, especially as this relates to their pairing and
rotational characteristics.
Further, the GXPF1 interaction is used to determine the strength parameter
of a quadrupole term that can be used to augment an isovector-pairing model
interaction with $\mathrm{Sp}(4)$ dynamical symmetry, which in turn is shown
to yield reasonable agreement with the low-lying energy
spectra of $^{58}$Ni and $^{58}$Cu.
\end{abstract}

\begin{keyword}
comparison of interactions, isovector pairing, quadrupole features, 
effective interactions, realistic
potentials, spectral distribution theory, Sp(4) dynamical symmetry
\PACS{21.60.Fw\sep 21.30.Fe\sep 21.10.Re\sep 21.60.Cs}
\end{keyword}
\end{frontmatter}


\section{Introduction}
Realistic $NN$ potentials, whether derived from meson exchange theory
(e.g.,\cite{MachleidtSS96M01}) or chiral effective field theory
(e.g.,\cite{EntemM03}), and their effective interaction
derivatives, provide no $a~priori$ indication regarding how well they
may or may not reproduce prominent features of nuclei, such as pairing
gaps in nuclear energy spectra or enhanced electric quadrupole transitions
in collective rotational bands, until actually employed in shell-model
calculations. While such calculations are often laborious and model
dependent, a simple and straightforward evaluation of an interaction can
be made using spectral distribution theory \cite{FrenchR71,ChangFT71}.
Indeed, spectral distribution methods can yield a deeper understanding of
the nature of an interaction and above all, its role in the development
of collective and correlated many-nucleon motion
\cite{Draayer73,HechtDraayer74,HalemaneKD78,Kota79,Rosensteel80,KotaPP80,CounteeDHK81,DraayerR82}.
In particular, spectral distribution theory can be used to show
through correlation coefficient measures the similarity of interactions.
Such analyses are independent of the averages of the interactions and
yield an overall comparison across a broad domain of nuclei beyond what
can be achieved by overlaps of nuclear states or detailed comparisons
of two-body interaction matrix elements.

In this paper we examine three modern $fp$-shell interactions, 
specifically,
two interactions
denoted as \CDB~  and
\CDBt~\cite{PopescuSVN05} based on the CD-Bonn realistic potential
\cite{MachleidtSS96M01} as well as \Gm~\cite{HonmaOBM04}. The \Gm~ 
effective interaction is obtained
from a realistic G-matrix interaction based  on the Bonn-C potential 
\cite{Gint} by  adding empirical
corrections determined through systematic fitting to experimental 
energy data in the
$fp$ shell. The CD-Bonn potential is a charge-dependent 
one-boson-exchange nucleon-nucleon ($NN$)
interaction that is
one of  the most accurate in reproducing the available proton-proton 
and neutron-proton scattering
data. Specifically, we use two-body matrix elements of an 
effective \CDB~interaction derived
from the CD-Bonn
potential for $0\hbar\Omega $ no-core shell model (NCSM) 
calculations in the $fp$ shell.  In addition, the
\CDBt~interaction introduces phenomenological isospin-dependent 
central terms plus a tensor force
with strengths  and ranges determined in $0\hbar\Omega $ NCSM 
calculations to achieve
an  improved description of the $A=48$ Ca, Sc and Ti isobars.
In the regard, we use spectral distribution theory to 
provide an assessment of
differences between the  novel \CDBt~ interaction and \CDB as well as 
a comparison of these interactions
with \Gm,
which has been shown to reproduce nuclear energy spectra throughout 
the $fp$ shell
\cite{HonmaOBM04}. 

The likely success of the three interactions for 
reproducing pairing and rotational
spectral features is examined in a comparison to a 
pairing-plus-quadrupole model interaction, which
combines a \Spn{4} dynamically symmetric model interaction 
\cite{SGD03stg,SGD04} for description
of like-particle and proton-neutron (isovector) pairing correlations 
with a SU(3) symmetric term
that governs a shape-determined dynamics. If the model and effective 
interactions  are
strongly correlated, then the latter will reflect the characteristic 
properties of the
simpler model Hamiltonian, which in turn may be used as a good approximation.

The present study, which is complementary to a similar \flevel
analysis \cite{SDV06}, focuses on the
\upfp-shell domain,
which includes neutron-deficient and $N\approx Z$ nuclei along 
the nucleosynthesis
$rp$-path and unstable nuclei currently explored in radioactive beam 
experiments
\cite{Langanke98,Hosmer05}. The analysis of the \upfp~results reveals 
overall properties of the
interactions under consideration different from those observed in the 
\flevel orbit.

Several detailed reviews of  the nuclear shell model and its 
applications have been published recently
\cite{CaurierMNPZ05,Brown01,OtsukaHMSU01} that delve into related key 
physics issues that are
explored in the present work.  However, the spectral distribution 
analysis provided here is novel and
sheds considerable light on the features of new $fp$-shell 
interactions, some of which have been
developed since those reviews were completed.

\section{Theoretical Framework}

The theory of spectral distributions is an excellent approach for 
studying microscopic interactions
\cite{ChangFT71,DraayerOP75,Potbhare77} and continues to be a 
powerful concept  with recent
applications in quantum chaos, nuclear reactions and
nuclear astrophysics including studies on nuclear
level densities, transition strength densities, and
parity/time-reversal violation
(for example, see
\cite{FrenchKPT88,KotaM94,TomsovicJHB00,GomezKKMR03,HoroiGZ04and03,Li03,ChavdaPK04,Kota05,AngomK05,Zhao05}).
The significance of the method is related to the fact that low-order
energy moments over a
certain domain of single-particle states, such as the energy
centroid of an interaction
(its average expectation value) and the deviation from  that average,
yield valuable information about the interaction that is of
fundamental importance
\cite{HalemaneKD78,CounteeDHK81,Potbhare77,Ratcliff71,French72,DaltonBV79,French83,DraayerR83a,SarkarKK87}
without the need for carrying out large-dimensional matrix diagonalization
and with little to no limitations due to the dimensionality of the
vector space.
Note that if one were to include higher-order energy moments, one
would obtain more detailed results that, in principle, should eventually
reproduce those of conventional microscopic calculations.

Spectral distribution theory (see Appendix for basic mathematical
definitions and notation
introduction) combines important features, the most significant of
which are as follows:
\begin{enumerate}
\item The theory provides a precise measure, namely, the correlation
coefficient, for the
overall similarity of two interactions.  Literally the correlation
coefficient is a measure of the
extent to which two interactions ``look like" (are correlated with)
one another. In this respect,
correlation coefficients can be used to extract information how well
pairing/rotational features are
developed in interactions, which may differ substantially
from an individual comparison of
pairing/quadrupole interaction strengths
\cite{SDV06}.

\item It gives an exact prescription for identifying the {\it pure} zero-
(centroid), one- and two-body parts of an interaction under a given
space partitioning. Therefore,
major properties follow:
\begin{enumerate}
      \item The correlation coefficients are independent of the
interaction centroids. (A direct
comparison of two-body matrix elements provided by $NN$ potentials
may be misleading,
especially when the averages of the interactions differ considerably.)

      \item  The pure one-body part of an interaction, the so-called induced
single-particle energies (Eq. \ref{ispeTS}), is naturally 
identified in the framework of
spectral distribution theory and is indeed the average monopole 
interaction (compare to
\cite{HonmaOBM04}). As such it influences the evolution of the shell structure,
shell gaps and binding energies \cite{Otsuka01}.

      \item  The pure two-body part is essential for studies of detailed
property-defining two-body  interactions beyond strong mean-field effects.
\end{enumerate}

\item The correlation coefficient concept can be propagated
straightforwardly beyond the defining
two-nucleon system to derivative systems with larger numbers of nucleons
\cite{ChangFT71} and higher values of isospin \cite{HechtDraayer74}.
This, in addition to
the two-nucleon information provided by alternative approaches
(e.g.,\cite{DufourZ96}), yields
valuable overall information, without a need for carrying out
extensive shell-model calculations,
about the universal properties of a two-body interaction in shaping
many-particle nuclear systems.

\end{enumerate}

Group theory underpins spectral distribution theory
\cite{FrenchR71,ChangFT71,HechtDraayer74,Kota79,Parikh78}. The model space is
partitioned according to particular group symmetries and each
subsequent subgroup partitioning yields finer and more detailed
spectral estimates. Specifically,
for $n$ particles distributed over $4\Omega$ single-particle
states, the spectral
distribution averaged over all $n$-particle states associated with
the \Un{}{4\Omega} group
structure is called a {\it scalar} distribution (denoted by ``$n$" in
the formulae)
and the spectral distribution averaged over the ensemble of all
$n$-particle states of isospin $T$ associated with $\Un{}{2\Omega}
\otimes \Un{}{2}_T$ is called an {\it isospin-scalar} distribution
(denoted by ``$n,T$").

For a spectral distribution $\alpha $ ($\alpha $ is $n$ or $n,T$), the
correlation coefficient between two Hamiltonian operators,
$H$ and $H^\prime$, is defined as
\begin{equation}
\zeta ^\alpha _{H,H^\prime }=\frac{\langle (H^\dagger -\langle
H^\dagger \rangle
^\alpha ) (H^\prime-\langle H^\prime \rangle ^\alpha )
\rangle ^\alpha }{\sigma _H \sigma _{H^\prime}}
=\frac{\langle H^\dagger H^\prime\rangle ^\alpha -\langle H^\dagger \rangle
^\alpha \langle H^\prime \rangle ^\alpha }{\sigma _H \sigma _{H^\prime}},
\label{zeta}
\end{equation}
where the ``width" of $H$ is the positive square root of the
variance,
\begin{equation}
(\sigma ^\alpha _{H})^2=\langle (H-\langle H\rangle ^\alpha )^2
\rangle ^\alpha
=\langle H^2 \rangle ^\alpha -(\langle H \rangle ^\alpha )^2,
\label{sigma}
\end{equation}
and the steps for computing these quantities are outlined in the
Appendix. The average values, $\langle \hat O \rangle ^\alpha$, related
to the trace of an operator $\hat O$ divided by the dimensionality of the
space, are given in
terms of the ensemble considered.
In the (isospin-)scalar case, the correlation will be denoted by $\zeta ^n$
($\zeta ^{n,T}$) or simply $\zeta $ ($\zeta ^T$) for $n=2$.
The significance
of a positive correlation coefficient is given by Cohen \cite{Cohen88_03} and
later revised to the following table:
\begin{table} [h]
\caption{Interpretation of a correlation coefficient. \label{tab:cc}}
\smallskip
\begin{small}\centering
\begin{tabular*}{\textwidth}{@{\extracolsep{\fill}}rrrrrrrrrr}
\hline  \noalign {\smallskip}
trivial  & small     & medium   & large    & very large & nearly perfect &
perfect \\
0.00-0.09 & 0.10-0.29 & 0.30-0.49 & 0.50-0.69 & 0.70-0.89   & 0.90-0.99      &
1.00\\
\hline
\end{tabular*}
\end{small}
\end{table}

  From a geometrical perspective, in spectral distribution theory 
every interaction is associated with a
vector and the correlation coefficient $\zeta $ (Eq. \ref{zeta}) 
defines the angle (via a normalized
scalar product) between two vectors of length $\sigma $ (Eq. 
\ref{sigma}). Hence, $\zeta _{H,H^\prime}$
gives the normalized projection of $H$ onto the $H^\prime$ 
interaction (or $H^\prime$ onto $H$). In
addition, $(\zeta _{H,H^\prime})^2$ gives the percentage of $H$ that 
reflects the characteristic
properties of the $H^\prime$ interaction.

The pairing and rotational characteristics of an  
interaction can be probed through its
projection onto a model microscopic Hamiltonian that describes 
isovector pairing correlations and
collective rotational excitations. While the latter possess a clear 
SU(3) symmetry \cite{Elliott}
within the framework of the harmonic oscillator shell model, the 
former  respect a
\Spn{4} dynamical symmetry \cite{Hecht,EngelLV96,KanekoHZPRC59}. 
Specifically, we
employ the pairing-plus-quadrupole model interaction
\begin{equation}
H_M=H_{\spn{4}}+\HQ,\
H_Q= -\frac{\chi }{2} Q \cdot Q ,
\label{HM}
\end{equation}
where \Hsp~ is an isoscalar \Spn{4}-dynamically symmetric interaction \cite{SGD04}
for a system of $n$ valence
nucleons (an eigenvalue of $\hat N$) in a $4\Omega
$-dimensional space,
\begin{equation}
\Hsp =-G
\sum _{i=-1}^{1}
\textstyle{
\hat{A}^{\dagger }_{i}
\hat{A}_{i}-\frac{E}{2\Omega} (\hat{T}
^2-\frac{3\hat{N}}{4 })
-C\frac{\hat{N}(\hat{N}-1)}{2}-\epsilon
\hat{N},
}
\label{clH}
\end{equation}
with two-body antisymmetric $JT$-coupled matrix elements (Appendix Eq. \ref{V2ndQF}) for
$\{r\le (s,t);\ t\le u\}$ orbits,
\begin{equation}
W_{rstu}^{JT}=-G_0
\textstyle{
\frac{\sqrt{\Omega _r \Omega _t}}{\Omega }
}
\delta_{J0}\delta_{T1}
\delta_{rs}\delta_{tu} -\{-E_0[(-)^T+\half]+C\}\delta_{rt}\delta_{su},
\label{W0me}
\end{equation}
where $ G_0=G+\frac{F}{3},\ E_0=(\frac{E}{2\Omega}+\frac{D}{3})$,
$G,F,E,D$ and $C$ are interaction strength parameters and $\epsilon >0$ is the
Fermi level energy (see Table I in Ref.\cite{SGD04} for parameter estimates).
The \spn{4} algebraic structure is exactly the one needed in nuclear
\IASs~ to describe
proton-proton ($pp$), neutron-neutron $(nn)$ and proton-neutron
($pn$) isovector pairing
correlations (accounted by the pair annihilation (creation)
operators $\hat{A}_{+1,-1,0}^{(\dagger)}$) and isospin symmetry. The latter is reflected by
the isospin operator
$\hat{T}^2$ and related to a $J$-independent isoscalar ($T=0$) $pn$
force.
The most general model interaction with \Spn{4}  dynamical symmetry 
\cite{SGD04} has been found
to provide for a reasonable microscopic description of the 
pairing-governed \IASs~ in light and medium
mass nuclei and to account quite well for the observed detailed 
structure  beyond mean-field
effects such as the $N=Z$ anomalies, isovector pairing gaps and 
staggering effects \cite{SGD03stg}.
In addition, the \Hsp~ interaction (\ref{clH}) has been shown to 
strongly correlate, especially when the
quadrupole term is introduced, with the \CDB, \CDBt~and \Gm~ 
interactions in the \flevel orbit
\cite{SDV06}.

The \HQ-term in (Eq. \ref{HM}) is the part of the pure two-body 
$H_Q(2)$ quadrupole-quadrupole
interaction that, in the vector
algebra terminology, is
orthogonal to the pure two-body \Spn{4} Hamiltonian
\cite{HalemaneKD78}. This is because the \Spn{4} interaction
contains a part of the quadrupole-quadrupole interaction that is not
negligible as revealed by the correlation between $H_Q$ and \Hsp.
Namely, in the scalar
case  it is 15\% ($\flevel$), 29\% ($\fFive $)
and 29\% ($\plevels$), and for the T=1 part of the interactions, it is
34\% ($\flevel$), 58\% ($\fFive$) and 58\% ($\plevels$).

Such a Hamiltonian (Eq. \ref{HM}) does not affect the centroid of \Hsp~ because
$\HQ$ is traceless. In this way this collective interaction
preserves the
shell structure that is built into \Hsp~ and established by a
harmonic oscillator
potential and as a result is favored in many studies
\cite{HalemaneKD78,CounteeDHK81}.

\section{Results and Discussions }

The similarity of the \CDB, \CDBt\, and \Gm~ interactions, which will 
be denoted as $H_0$, and their
pairing/rotational characteristics can be tracked in many-nucleon systems
\cite{KotaPP80} through the propagation
formulae (Appendix Eqs. \ref{<HH'>n}, \ref{<HH'>nT}). The latter
determine how the  averages
extracted from the two-nucleon matrix elements get carried forward
into many-nucleon systems. This
propagation of information is model-independent.

The present investigation focuses on the \upfp-shell domain  and
is complementary to a similar \flevel analysis \cite{SDV06}  (a few
results from that study are presented in Section \ref{sec:singlej} 
for completeness). Such a
partitioning of the $fp$ oscillator shell follows naturally from a
splitting of these two regions
by a strong spin-orbit interaction.
We examine the $H_0(2)$ pure two-body part of the $fp$-shell 
interactions and how it is
correlated to the model interaction (Eq. \ref{HM}). The latter, in addition
to its centroid, is pure two-body in the \upfp~model space
because of the assumption for \Hsp~of constant Fermi level energy and 
fixed interaction strengths
throughout the entire region. The
significance of the correlation coefficients between pure two-body
interactions \cite{Potbhare77}
reflect the fact that nuclear states, their collective properties and
configuration mixing, are
solely shaped by the pure two-body part of an interaction, while the
one-body part, albeit of a
considerable significance, trivially reorders the states in the
nuclear energy spectrum. In
addition, such analyses are free of the one-body influence including
induced single-particle
energies, which are related to the monopole interaction
\cite{DufourZ96,Otsuka01,HonmaOBM04}, and
external single-particle energies. The latter are introduced when a
core is assumed, as for the $0\hbar
\Omega$ $^{40}$Ca-core shell model using the \Gm~ interaction. For the
$0\hbar \Omega$ NCSM calculations with \CDB~or \CDBt, the two-body 
matrix elements
specifying the particle-core interactions supplant the role of external
single-particle energies.  These additional two-body matrix elements 
together with the external
single-particle energies for \Gm~ are not included in the present analyses.

In our study, we vary only $\chi$, the quadrupole strength parameter in (Eq.
\ref{HM}), to find its optimal value (which is an exact  solution) by
maximizing the $\zeta$
correlation coefficient \cite{Chang78} between the model $H_M$
interaction and the pure two-body
part $H_0(2)$ of each of the effective interactions under
consideration. We do not alter the
parameters of the \Spn{4} model, which have already been shown in an
appropriate domain of states to
be valid for reproducing various quantities (such as binding energies
and pairing gaps) and are in
agreement with estimates available in literature \cite{SGD03stg,SGD04}.

\begin{figure}[e]
\centerline{\epsfxsize=3.5in\epsfbox{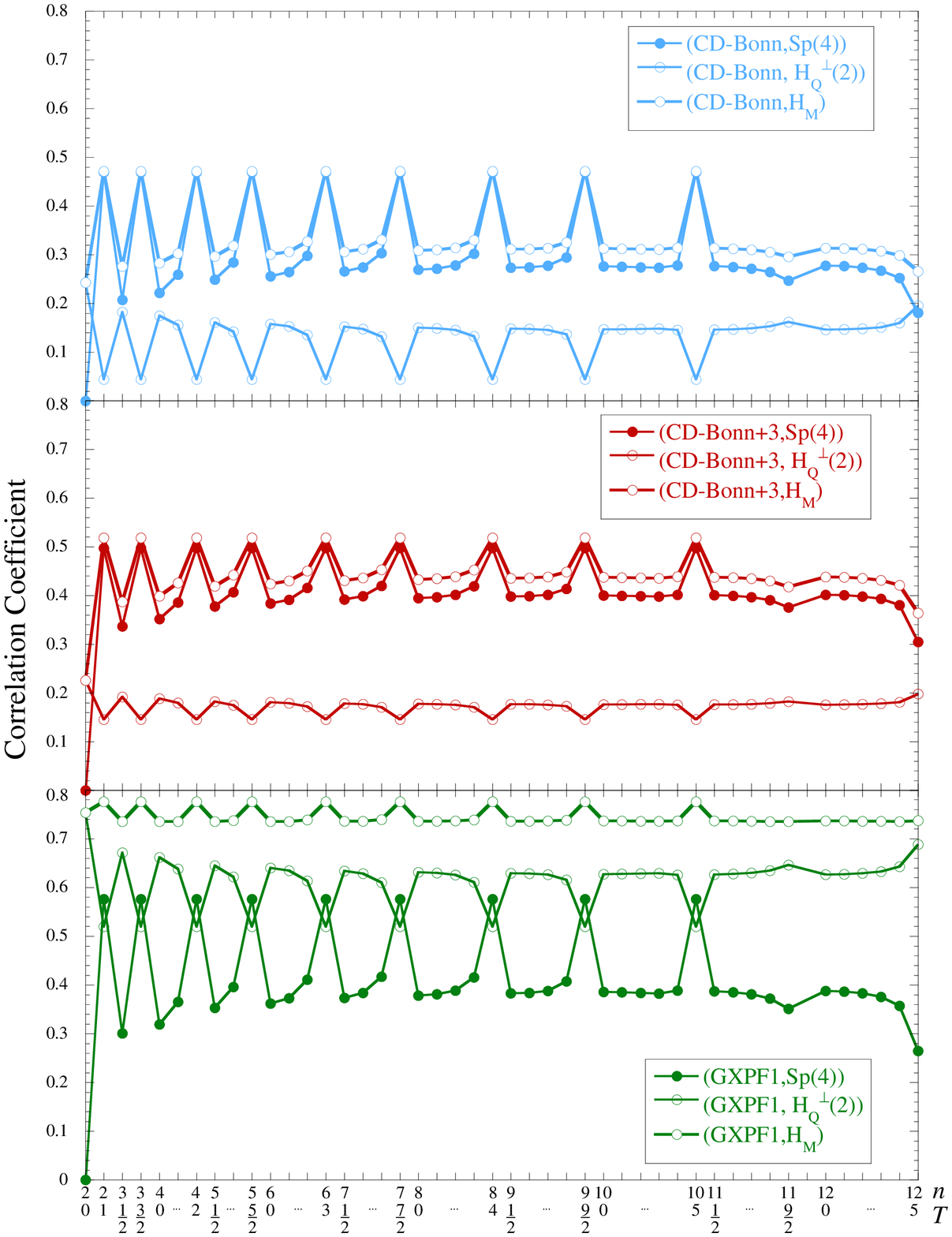}
}
\caption{Correlation coefficients of the pure two-body
\CDB~(blue), \CDBt~(red) and \Gm~(green) interactions with \Hsp~ (filled
symbols), \HQ~(transparent symbols) and $H_M$ (empty symbols) in the
\upfp~ shell for the isospin-scalar distribution. For each valence
particle number, $n$, the isospin $T$ varies as
$\frac{n}{2},\frac{n}{2}-1,\dots, 0(\half)$. The figures are symmetric
with respect to the sign of $n-2\Omega$ ($\Omega=6$).}
\label{upfp}
\end{figure}

In both scalar (Table \ref{tab:upfpS}) and isospin-scalar (Fig.
\ref{upfp}) distributions, the
$\zeta^{n(T)}_{H_{0}(2),H_M}$ correlation with the pairing+quadrupole 
$H_M$ interaction is
stronger for GXPF1 compared to \CDB~and \CDBt.
Hence, other types
of interactions
that do not correlate with the pairing and quadrupole-quadrupole
interactions (of fixed strength
throughout the \upfp~shell) comprise a relatively small part of the
pure two-body \Gm~ interaction.
They are weakest for the
$T=n/2$ group of states (Fig. \ref{upfp}) for all the three 
interactions.  In addition, the results
reveal that the symplectic \Spn{4} dynamical symmetry of \HM,
especially when compared to both
CD-Bonn interactions and in the highest-isospin ($T=n/2$) group 
of states, is only
slightly broken by $\HQ$ (Table
\ref{tab:upfpS}, fifth row and Fig.
\ref{upfp}). The \Spn{4} symmetry breaking is related to the
correlation coefficient
of \HQ~ with $H_M$ or equivalently to the ratio of their norms
\cite{French67}, where $(\zeta _{H_M,\HQ})^2=1-(\zeta _{H_M,\Hsp})^2=1-R_{\Spn{4}}$
(Table \ref{tab:upfpS}).

\begin{table}[e]
\caption{Scalar distribution correlation coefficients for many-nucleon systems of the $H_0(2)$
pure two-body part of the \CDB, \CDBt~and \Gm~ interactions  with \HM~(Eq.
\ref{HM}), \Hsp~(Eq. \ref{clH}), \HQ, and with the pure two-body full quadrupole-quadrupole 
interaction, $H_Q(2)$. The $R_{\Spn{4}}=(\zeta
_{H_M,\Hsp})^2$ quantity gives the part, in
\%, of $H_M$ that is \Spn{4} symmetric.}
\center{
\begin{tabular}{lcccc}
\hline
& \CDB    & \CDBt                               &   \Gm\\
\hline \hline
$\zeta _{H_0(2),H_M}$     & 0.57    &  0.54    &  0.83  \\
$\zeta _{H_0(2),\Hsp}$    & 0.55    &  0.50    &  0.65  \\
$\zeta _{H_0(2),\HQ}$     & 0.14    &  0.20    &  0.51  \\
$\zeta _{H_0(2),H_Q(2)}$  & 0.28    &  0.33    &  0.67  \\
$R_{\Spn{4}}$             & 93.1\%  &  85.7\%  &  61.3\%  \\
\hline
\end{tabular}
}
\label{tab:upfpS}
\end{table}

Two correlation coefficients, which in the present study are 
independent of any
interaction strength parameters, are of particular interest. 
Specifically, the isospin-scalar space
partitioning is where the ability of an interaction to form 
correlated pairs and hence
reproduce prominent pairing gaps is detected via 
$\zeta_{H_0(2),\Hsp}^{n,T}$.  The capability of
an interaction to describe rotational collective motion, and hence to 
reproduce rotational bands and
enhanced electric quadrupole transitions, can be detected via its 
correlation to the full $H_Q(2)$
quadrupole-quadrupole two-body interaction, $\zeta_{H_0(2),H_Q(2)}^{n(T)}$.

\begin{figure}[th]
\centerline{\epsfxsize=3.5in\epsfbox{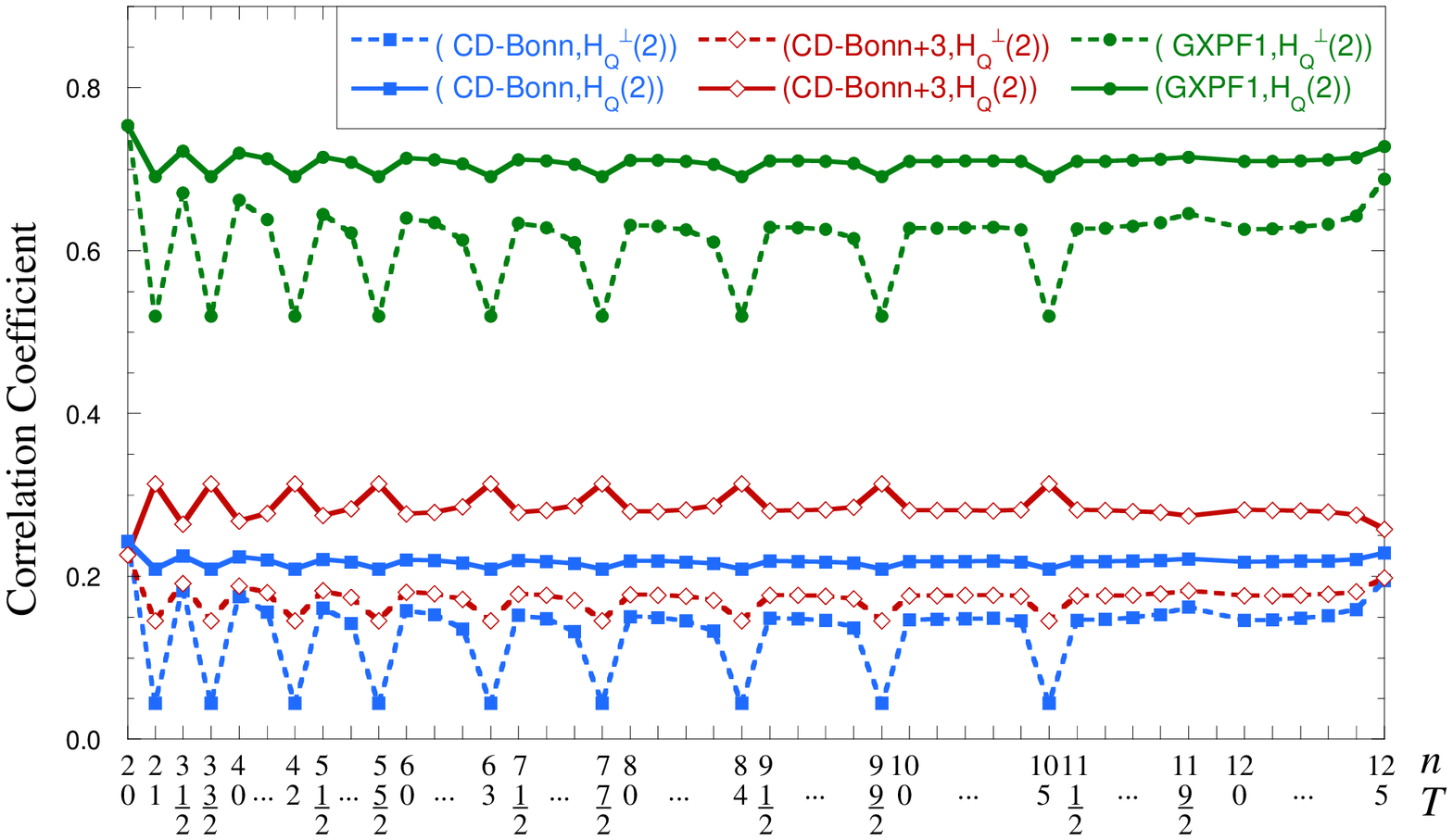}}
\caption{Comparison between the orthogonal  \HQ~ and the full two-body $H_Q(2)$
quadrupole-quadrupole interactions in their correlation to the pure 
two-body part of the
$fp$-shell interactions, \CDB~ (blue squares), \CDBt~(red diamonds) and
\Gm~(green circles), in the
isospin-scalar distribution. For each valence particle number, $n$,
the isospin $T$ varies as
$\frac{n}{2},\frac{n}{2}-1,\dots, 0(\half)$. The figure is symmetric with
respect to the sign of $n-2\Omega$ ($\Omega=6$).}
\label{QLcmpQ}
\end{figure}

The results show a large correlation with isovector pairing of 
the $T=1$ part of the $fp$-shell
interactions under considerations ($n=2$), especially for \Gm, and a 
good tendency towards development
of pairing correlations in the $T=n/2$ states (Fig. \ref{upfp}).  In 
both scalar and isospin-scalar
cases (Table
\ref{tab:upfpS}, fourth row and Fig.
\ref{QLcmpQ}), the rotational features are more fully developed for 
\Gm~ and less for
\CDBt~ and
\CDB~ (Fig. \ref{QLcmpQ}) different from the outcome in the \flevel 
region especially for the
$T=1$ part of \CDB.

In short, the \Gm~ interaction is expected to reproduce spectral
features like pairing gaps and rotational bands observed in the
\upfp~ nuclei, while it is unlikely for both CD-Bonn
interactions to fully reflect the rotational properties of these nuclei. In
comparison, the \CDBt~ interaction in the \flevel orbit exhibits
well-developed pairing and rotational characteristics \cite{SDV06}.
Such a difference in the behavior of \CDBt~ within both
regions, \flevel~ and \upfp, may reflect the fact that this interaction was
determined through a reproduction of the low-lying energy spectrum 
and binding energy
of $A=48$ \flevel nuclei.

The different extent to which the \Gm~ interaction compared to the
\CDB~and \CDBt~ interactions
reflects development of pairing correlations and collective
rotational modes in the \upfp~ domain
may be the reason why their pure two-body part do not correlate
strongly as, for example, \CDB~and
\CDBt~ do. Namely, in the (isospin-) scalar case the pure two-body correlations\footnote{In
the isospin-scalar case, the
correlations vary slightly with the particle number and isospin and
average values are quoted.} are 0.90 (0.88) between \CDB~ and
\CDBt~ and only 0.56 (0.37) between \CDB~ and \Gm~ and 0.53 (0.40) 
between \CDBt~ and \Gm. In addition,
the isospin-scalar correlation coefficients involving the significant 
induced pure one-body  (monopole)
contribution, $\lambda _j^{T}$ (Appendix Eq. \ref{ispeTS}), differ 
between \Gm~ and the two \CDB~
interactions. Their behavior, especially below mid-shell, reflects 
the similarity of the corresponding
$T=0$ induced  single-particle energies and the opposite signs of the 
corresponding $\lambda
_{2p_{3/2}}^{T=1}$ and  $\lambda _{\fFive}^{T=1}$. 

In summary, when compared to the original \CDB, $0\hbar \Omega$ 
NCSM calculations with \CDBt~for
\upfp~ nuclei are likely to achieve an improved description of
many-body spectral phenomena, associated with pair formation 
(especially when  $T\ne n/2$) and  with
rotational motion (especially in the  highest-isospin states). While 
such results lie slightly
closer to what is achieved by the \Gm~ interaction with two-body 
matrix elements directly adjusted
to experimental $fp$-shell data, larger NSCM model spaces for both 
CD-Bonn interactions  may be
necessary for a better reproduction of pairing and rotational 
spectral features throughout the \upfp~
domain. 

\subsection{Individual-Orbit Analysis \label{sec:singlej}}

One can further perform a partitioning of the $fp$-space to
single-$j$ orbits, $\flevel$,
$\fFive$, $2p_{1/2}$ and $2p_{3/2}$, to provide for more detailed
spectral measures that may
reflect  important fine effects that are otherwise averaged out when
the entire $fp$ major
shell is  taken into account. We have already illustrated such an
example and its
significance by exploration of the \flevel orbit \cite{SDV06}. 
Individual orbit analyses
render correlation
coefficients that are free of the influence of any one-body
interaction [by definition, (Appendix
Eq. \ref{espe1},\ref{ispeS},\ref{ispeTS})]. However, due to the small
model space, the
$2p_{1/2}$ and $2p_{3/2}$ orbits are combined and in their joint region
only the pure two-body part of the interactions is considered.

In the scalar distribution, a good portion, 53\% to 98\%, of the 
\HM~pairing+quadrupole
model interaction is described solely by the \Hsp~
interaction  ($R_{\Spn{4}}$, Table \ref{tab:jS}).
The $H_0$ $fp$-shell
interactions exhibit a quite well-developed rotational character
(Table \ref{tab:jS},
$\zeta _{H_0,H_Q}$) except for \CDB~and \CDBt~ in the \plevels
region. Besides these cases, the
\HM~ model interaction, as revealed by $\zeta _{H_0,H_M}$ in Table
\ref{tab:jS}, can be used as a very good
approximation to the $fp$-shell interactions within each of the 
domains considered.
\begin{table}[th]
\caption{Correlation coefficients for many-nucleon systems of the 
\CDB, \CDBt~ and \Gm~
interactions, $H_0$, with \HM (Eq. \ref{HM}),
\Hsp (Eq. \ref{clH}), and the full
quadrupole-quadrupole interaction, $H_Q$. The $R_{\Spn{4}}=(\zeta
_{H_M,\Hsp})^2$ quantity gives the part, in
\%, of $H_M$ that is \Spn{4} symmetric.}
\center{
\begin{tabular}{llcccc}
\hline
\multicolumn{5}{c}{Scalar Distribution} \\
    & & $1f_{7/2}$    & $1f_{5/2}$
                                  &   $2p_{1/2}2p_{3/2}$$^a$\\
\hline \hline
\CDB      &   $\zeta _{H_0,H_M}$     & 0.81    &  0.76    &  0.61  \\
          &   $\zeta _{H_0,\Hsp}$    & 0.66    &  0.70    &  0.58  \\           
           &   $\zeta _{H_0,H_Q}$  & 0.69    &  0.64    &  0.17  \\
           &   $R_{\Spn{4}}$         & 65.9\%  &  83.7\%  &  91.7\%  \\
\CDBt     &   $\zeta _{H_0,H_M}$     & 0.87    &  0.86    &  0.52  \\
          &   $\zeta _{H_0,\Hsp}$    & 0.64    &  0.67    &  0.51  \\           
           &   $\zeta _{H_0,H_Q}$  & 0.80    &  0.82    &  0.22  \\
           &   $R_{\Spn{4}}$         & 53.4\%  &  60.6\%  &  98.3\%  \\
\Gm       &   $\zeta _{H_0,H_M}$     & 0.93    &  0.84    &  0.90  \\
           &   $\zeta _{H_0,\Hsp}$    & 0.76    &  0.77    &  0.70  \\          
           &   $\zeta _{H_0,H_Q}$  & 0.78    &  0.69    &  0.85   \\
           &   $R_{\Spn{4}}$        & 67.9\%  &  84.7\%  &  61.7\%  \\
\hline
\multicolumn{5}{l}{\footnotesize $^a$pure two-body part of the interactions}
\end{tabular}
}
\label{tab:jS}
\end{table}

The $\zeta _{H_0,\Hsp}^{T=1}$ correlation coefficients (Table \ref{tab:jTS})
show large $J=0$ isovector
coherence within each single-$j$ shell, particularly for the $T=1$ 
part of \CDBt~ and \Gm,
which are expected to
describe quite well phenomena of a pairing character, while for \CDB~ other
types of interaction compete with pair formation. The latter are of
the \HQ~ type for the
\flevel~and \fFive~ orbits, where the residual interactions are negligible.
In short, the simple \Spn{4} model interaction and especially its extended
pairing+quadrupole \HM~ interaction can reproduce reasonably well the
$T=1$ part of the three effective interactions under 
consideration within the orbits specified in
Table
\ref{tab:jTS}.
\begin{table}[th]
\caption{Correlation coefficients for a two-nucleon system of
the \CDB, \CDBt~ and \Gm~ interactions, $H_0$, with \HM (Eq. \ref{HM})
and \Hsp (Eq. \ref{clH}). $R_{\Spn{4}}=(\zeta 
_{H_M,\Hsp})^2$ gives the part, in \%, of $H_M$
that is \Spn{4} symmetric.}
\center{
\begin{tabular}{llcccc}
\hline
\multicolumn{5}{c}{Isospin-scalar Distribution, $T=1$} \\
    & & $1f_{7/2}$    & $1f_{5/2}$
                                  &   $2p_{1/2}2p_{3/2}$$^a$\\
\hline \hline

\CDB     &  $\zeta _{H_0,H_M}^{T=1} $  &0.95     &  1.00     &  0.59  \\
         &  $\zeta _{H_0,\Hsp}^{T=1} $ &0.61     &  0.57     &  0.54  \\           
           &  $R_{\Spn{4}}$    &41.5\%   & 32.3\%    &  84.2\%    \\

CD-Bonn     &  $\zeta _{H_0,H_M}^{T=1} $  &0.98     &  1.00     &  0.61  \\
+3terms   &  $\zeta _{H_0,\Hsp}^{T=1} $ &0.85     &  0.93     &  0.59  \\
           &  $R_{\Spn{4}}$    &73.9\%   &  86.9\%   &  92.1\%  \\

\Gm     &  $\zeta _{H_0,H_M}^{T=1} $  &0.96     &  1.00     &  0.94  \\
       &  $\zeta _{H_0,\Hsp}^{T=1} $ &0.71     &  0.86     &  0.69\\           
           &  $R_{\Spn{4}}$    &54.5\%   &  74.0\%   &  53.5\%  \\
\hline
\multicolumn{5}{l}{\footnotesize $^a$pure two-body part of the interactions}
\end{tabular}
}
\label{tab:jTS}
\end{table}

Within individual orbits a very close similarity is observed between 
the effective and model
interactions as well as in nuclear systems with more than two 
nucleons. However, more prominent
differences among the interactions appear in the multi-$j$ \upfp~
domain especially concerning both \CDB~interactions. This may
indicate that the inter-orbit
interactions do not respect strongly the symmetries imposed in the
model interaction.
In addition, the interaction strengths may differ from one orbit to
another. While they do not
affect correlation coefficients in the singe-$j$ cases, their
relative strength is of a
great importance for multi-$j$ analysis.

\subsection{Energy Spectra for $A=58$ Nuclei}

The results presented above show that the $H_M$ model Hamiltonian
(Eq. \ref{HM}) can be used in
the \upfp ~ region as a quite good approximation of the pure two-body
part of the \Gm~
interaction (Table \ref{tab:upfpS} and Fig. \ref{upfp}).  This implies 
that both interactions are
expected to yield  energy spectra of a similar pattern.

As an illustrative example, we apply  the simpler model interaction (Eq.
\ref{HM}) to a nuclear system of
two nucleons in the \upfp~ region without any parameter variation.
Particularly, we assume a $^{56}$Ni-core and that both nucleons in
$^{58}$Ni and
$^{58}$Cu occupy the \upfp~ orbits with reasonable probability
\cite{HonmaOBM04}.  For a
description of the low-lying structure of these nuclei, external
single-particle energies are
needed to rescale at the end the eigenvalues of the model
Hamiltonian. This is performed trivially
due to the microscopic structure of the model eigenstates, which are
constructed in terms of
fermion creation operators. We adopt single-particle energies that
are derived from the
$^{57}$Ni energy spectrum. To a very reasonable degree, these
energies reflect the influence of the
\flevel orbit and the core mean-field contribution.

The model Hamiltonian uses a $\chi$ value of $0.027$ that we obtained
through a comparison of $\HM$ to
\Gm~ within a scalar distribution. The reason is that the pure
two-body \Gm~interaction holds the best
correlation coefficient to the model interaction in the \upfp~ region
(Table \ref{tab:upfpS}). In
addition, the energy spectra for $^{58}$Ni and $^{58}$Cu are found to
be closely reproduced by
shell model calculations with the \Gm~ interaction in the full $fp$ shell
\cite{HonmaOBM04}.

\begin{figure}[th]
\centerline{\epsfxsize=3.5in\epsfbox{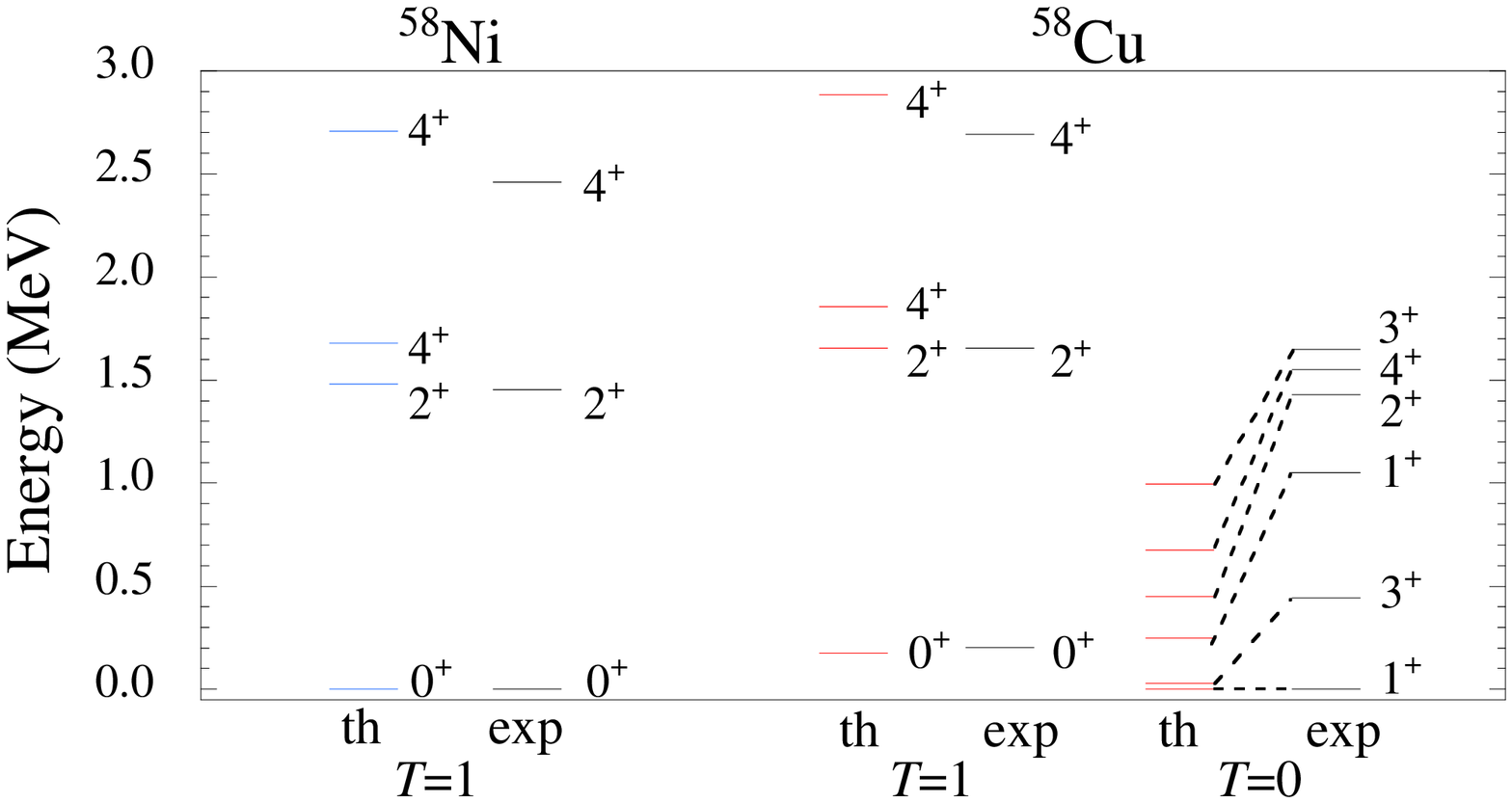}}
\caption{Theoretical (`th') low-lying energy spectra for $^{58}$Ni
(left, blue) and $^{58}$Cu (right,
red) compared to experiment (`exp', black). The theoretical
calculations are performed in the \fpg
major shell with the $H_M$ model interaction (Eq. \ref{HM} with
$\chi=0.027$) and with
single-particle energies derived from $^{57}$Ni experimental energy levels.}
\label{enSpectra}
\end{figure}

We extend the model space to include the
\glevel orbit as it intrudes in the \upfp~domain. This is exactly the
space where the \Spn{4} model was applied and interaction strength parameters
determined \cite{SGD04}.
The results (Fig. \ref{enSpectra}) show a very good reproduction of
the low-lying
$T=1$ spectra in $^{58}$Ni and $^{58}$Cu, especially the lowest $2^+$
states for both nuclei and
the first $0^+$ ($T=1$) state above the $^{58}$Cu ground state. Both
states are of particular
significance. On the one hand, the energy difference between the
lowest $0^+$ and $2^+$ states is
believed to be directly affected by the formation of correlated pairs
in the lowest $0^+$
state (ground state for even-even nuclei)  and the pairing gap that
occurs below the first excited
$2^+$ state of a broken pair. On the other hand, the
$0^+(T=1)$ to $1^+(T=0)$ energy difference in $N=Z$ odd-odd nuclei is
associated with the close interplay of
isovector (T=1, pairing correlations) and isoscalar (T=0)
interactions between protons and neutrons
in the same major shell. In addition, the
$T=0$ spectrum of $^{58}$Cu as predicted by the model interaction
possesses the same pattern of the
levels observed, namely, the $1^+$ ground state is followed by $3^+$,
$1^+$, $2^+$, $4^+$, and
$3^+$. The $^{58}$Cu $T=0$ spectrum appears narrower than the
experimental data, which
suggests that different quadrupole strengths for $T=0$ and $T=1$
need to be used.

In short, we demonstrate that strong correlations typically yield
very similar energy spectra and
reproduction of the overall pattern of the energy levels without any
adjustment of the
interaction strength parameters.

\section{Conclusions}

 In the present article, we have compared three modern 
$fp$-shell interactions, \CDB, \CDBt, and \Gm, based on realistic
nucleon-nucleon potentials, in the \upfp~ region by means of the 
theory of spectral distributions. We
focused on the weaker but property-defining two-body part of the 
interactions and studied their pairing
and rotational character in a comparison  to a pairing$+$quadrupole 
model interaction,  which includes
\Spn{4} dynamically symmetric isovector pairing correlations and a 
proton-neutron isoscalar force
together with a quadrupole-quadrupole interaction for description of
\SU{}{3} dynamically symmetric collective rotational mode.

The outcomes show that the pure two-body \Gm~ interaction, which is 
adjusted through systematic fitting
to experimental  energy data in the $fp$ shell, demonstrates well 
developed collective rotational
features and a tendency of its $T=1$ part towards formation of 
correlated pairs. In addition, it
strongly correlates with the pairing$+$quadrupole $H_M$ model 
interaction, which is reflected in the
quite good agreement between the experimental low-lying energy 
spectra of $^{58}$Ni and $^{58}$Cu
and the theoretical prediction based on $H_M$ in the \fpg~ major shell.
Individual-orbit analysis,
including the  $\flevel$, $\fFive$, $2p_{1/2}$, and $2p_{3/2}$ 
levels, shows considerably stronger
correlation to $H_M$ (up to $0.8-1.00$) and a clear pairing or/and 
rotational character for all the
three $fp$-shell interactions, which in addition correlate more 
strongly among themselves. In this
respect,  inter-orbit interactions may be a reason why both CD-Bonn 
interactions suitable for $0\hbar
\Omega$ NCSM $fp$-shell calculations invoke small to medium pure 
two-body correlations in the
\upfp~domain. While these correlations for \CDBt~ appear slightly 
superior to the ones for \CDB due to
a phenomenological correction, it may be interesting to investigate 
how they vary for effective
interactions based on the realistic CD-Bonn potential when 
higher-$\hbar \Omega$ configurations are
included in the no-core shell model analysis. 

In summary, based on these results, spectral
distribution theory appears to be a good framework for uncovering
fundamental  properties of realistic interactions and their 
effective interaction derivatives  in
many-nucleon systems. We find varying degrees of respect for selected 
underlying symmetries.  As
some of these symmetries have been demonstrated to be important for certain
spectral features, we have a tool for rapidly assessing the likely success
of these interactions for reproducing those spectral features.  For example,
it is unlikely that the \CDBt~ interaction will provide a fully satisfactory
description of the rotational properties of nuclei in the \upfp~ shell.
Given that this interaction was determined only with $A=48$ nuclear spectra
and binding energies \cite{PopescuSVN05}, future efforts at expanding the
region of its validity in the no-core shell model should benefit from the
analysis provided here.

\vskip 0.5cm
{\bf Acknowledgments}

One of us (JV) would like to thank Sorina Popescu, Sabin Stoica and
Geanina Negoita for valuable discussions.
This work was supported by the US National Science Foundation, Grant
Numbers 0140300 \& 0500291, and the Southeastern Universities Research
Association. This work was partly performed under the auspices
of the US Department of
Energy by the University of California, Lawrence Livermore
National Laboratory under contract No.  W-7405-Eng-48
and under the auspices of grants DE-FG02-87ER40371 \&
DE--AC02--76SF00515.

\vskip 0.5cm
{\bf Appendix}

The theory of spectral distributions (or statistical spectroscopy) is
well documented
in the literature \cite{FrenchR71,ChangFT71,French72,HechtDraayer74,Kota79}
and is accompanied
by computational codes \cite{Kota79,ChangDW82} for evaluating various
measures. The
purpose of this appendix is to specify the notation and ensure that
our definitions of
the summations and numerical factors that enter into such measures  are clearly
understood.

In standard second quantized form, a one- and two-body interaction
Hamiltonian is given
in terms of fermion creation $a_{ jm(1/2)\sigma }^\dagger =c_{ jm(1/2)\sigma
}^\dagger$ and annihilation
$a_{ j-m(1/2)-\sigma } = (-1)^{j-m+1/2 -\sigma }c_{ jm(1/2)\sigma }$
tensors, which create or annihilate a particle of type
$\sigma =\pm 1/2$ (proton/neutron) in a state of total angular momentum $j$
(half integer) with projection
$m$ in a finite space $2\Omega =\Sigma _j (2j+1)$,
\begin{eqnarray}
H&=&-\sum _{r\leq s}\sqrt{[r]} \varepsilon _{rs} \{a_r^\dagger
\otimes a_s\}^{(00)} \nonumber\\
&&-\frac{1}{4}\sum_{rstu \Gamma}
\sqrt{(1+\delta _{rs})(1+\delta _{tu})[\Gamma]} W_{rstu}^\Gamma
\{\{a_r^\dagger \otimes a_s^\dagger\}^\Gamma \otimes
\{a_t \otimes a_u\}^\Gamma \}^{(00)},
\label{V2ndQF}
\end{eqnarray}
where the labels are $r=\{j_r ,\tau _r= \half\}$,
$[r]=2(2j_r+1)$, and $[\Gamma]=(2J+1)(2T+1)$. In (Eq. \ref{V2ndQF}),
$\varepsilon _{rs}$ is the (external) single-particle energy  (hereafter we
consider no angular momentum degeneracy for two different radial quantum
numbers, $\varepsilon _{rs}=\varepsilon _r \delta_{rs}$) and
$W_{rstu}^{JT} =<rsJTMT_0|H|tuJTMT_0>$ is the two-body antisymmetric 
matrix element in the
$JT$-coupled scheme [$W_{rstu}^{\Gamma
}=-(-)^{r+s-\Gamma}W_{srtu}^{\Gamma }=
-(-)^{t+u-\Gamma } W_{rsut}^{\Gamma }=(-)^{r+s-t-u}W_{srut}^{\Gamma }=
W_{turs}^{\Gamma }$]. For an isospin nonconserving two-body
interaction of isospin
rank ${\mathcal T} $, the coupling of fermion operators is as follows,
$\{\{a_r^\dagger \otimes
a_s^\dagger\}^{JT}\otimes \{a_t \otimes a_u\}^{JT} \}^{(0{\mathcal T})}$, with
$W_{rstu}^{({\mathcal T}) J T}$ matrix elements.

      For a major shell that consists of several $s$ orbits, each of
degeneracy ${\mathcal N}_s$ (${\mathcal N}=\sum _s {\mathcal N}_s$),
the (traceless)
external single-particle energy of the $r^{\text{th}}$ orbit is obtained as
\begin{equation}
\tilde \varepsilon _r=\varepsilon _r- \varepsilon =\varepsilon _r-
\frac{1}{{\mathcal N}}\sum _s \varepsilon _s {\mathcal N}_s,
\label{espe1}
\end{equation}
where the average (external) single-particle energy is
$\varepsilon =\frac{1}{{\mathcal N}}\sum _s \varepsilon _s {\mathcal N}_s$.

{\bf Scalar Distribution.} For a two-particle system, the monopole
moment (centroid), which
is the average expectation value of the two-body interaction,  is
defined in the
scalar case as
\begin{equation}
W_c=\frac{1}{\binom{{\mathcal N}}{2}}\sum _{r \le s,
\Gamma}[\Gamma] W_{rsrs}^\Gamma
=\frac{\sum _{rs, \Gamma}[\Gamma] W_{rsrs}^\Gamma (1+\delta _{rs})}
{{\mathcal N}({\mathcal N}-1)},
\label{WcS}
\end{equation}
where ${\mathcal N}=4\Omega =2\sum _r(2j_r+1)$, the $\Gamma$-sum
goes over all
possible $(J,T)$ for given $r,s$, and $\binom{{\mathcal N}}{2}=\sum _{r \le s,
\Gamma}[\Gamma]$.
The traceless induced single-particle energy is constructed by
contraction of the
two-body interaction into an effective one-body operator under the
particular group
structure,
\begin{equation}
\lambda _r =\frac{1}{{\mathcal N}_r}\sum _{s,JT} [JT]
W_{rsrs}^{JT} (1+\delta _{rs})
-\frac{1}{{\mathcal N}} \sum _{tu,JT} [JT]
W_{tutu}^{JT} (1+\delta _{tu}).
\label{ispeS}
\end{equation}
For a system with one hole in the $r^{\text{th}}$ orbit, $\lambda _r$
corresponds to
the energy of a single particle as contributed by the interaction
with the valence
particles above the core.
In turn, the traceless pure two-body interaction is defined as
\begin{equation}
W_{rstu}^{JT}(2)=W_{rstu}^{JT}-(W_c +\frac{\lambda _r +\lambda
_s}{{\mathcal N}-2})
\delta _{rt}\delta _{su}.
\label{W2S}
\end{equation}

In order to calculate energy moments and their propagation for higher $n$
(and $T$) values, each interaction $H$ (consisting of one($k=1$)- and
two($k=2$)-body parts) needs to be expressed as a linear combination
of terms of
definite particle rank (irreducible tensors ${\mathcal H}_k(\nu )$ of rank $\nu
=0,1,2$), that is as a collection of pure zero-, one- and two-body
interactions. For
$n$ particles, the Hamiltonian can be rendered (the sum $\sum^*$ goes 
over $r\leq s$,
$t\leq u$ and $\Gamma=(J,T)$),
\begin{eqnarray}
H&=&n{\mathcal H}_1(0)+\binom{n}{2} {\mathcal H}_2(0)+{\mathcal
H}_1(1)+(n-1){\mathcal H}_2(1) +{\mathcal H}_2(2) \nonumber\\
&=&-n\varepsilon -
\binom{n}{2}W_c  -\sum _{r}[r]^{\half } (\tilde \varepsilon _{r}
+\frac{n-1}{{\mathcal N}-2}\lambda _{r})\{a_r^\dagger
\otimes a_r\}^{(00)}  \\
&&-
\textstyle{\sum ^*}
\frac{\sqrt{[\Gamma]}}{\sqrt{(1+\delta _{rs})(1+\delta _{tu})}}
W_{rstu}^\Gamma (2)\{\{a_r^\dagger \otimes a_s^\dagger\}^\Gamma \otimes
\{a_t \otimes a_u\}^\Gamma \}^{(00)},
\nonumber
\end{eqnarray}
for then the quantity that defines the correlation coefficient
(Eq. \ref{zeta}) is easily
computed for different particle numbers $n$,
\begin{eqnarray}
&\langle H^\dagger H^\prime\rangle ^n-\langle H^\dagger \rangle ^n\langle
H^\prime \rangle ^n=
\frac{n({\mathcal N}-n)}{{\mathcal N}({\mathcal N}-1)} \sum _{r} [\tilde
\varepsilon _{r}+
\frac{n-1}{{\mathcal N}-2}\lambda _{r}][\tilde \varepsilon ^\prime _{r}+
\frac{n-1}{{\mathcal N}-2}\lambda ^\prime _{r}] {\mathcal N}_r \nonumber \\
&+\frac{n(n-1)({\mathcal N}-n)({\mathcal N}-n-1)}
{{\mathcal N}({\mathcal N}-1)({\mathcal N}-2)({\mathcal N}-3)}
\textstyle{\sum ^*}
[\Gamma] W_{rstu}^\Gamma (2) {W^\prime }_{rstu}^\Gamma (2).
\label{<HH'>n}
\end{eqnarray}

{\bf Isospin-Scalar Distribution.}
Analogously, the centroid
is defined as,
\begin{equation}
W_c^T=\frac{2}{{\mathcal N} ({\mathcal N}+(-1)^T)}\sum _{r \le s, J}[J]
W_{rsrs}^{JT}
\label{WcTS}
\end{equation}
where ${\mathcal N}=2\Omega $. The $\lambda _r ^T$ traceless induced
single-particle
energy for orbit $r$ and the $W_{rstu}^{JT}(2)$ traceless pure
two-body interaction \cite{Kota79}
are defined as,
\begin{eqnarray}
\lambda _r ^T=\frac{1}{{\mathcal N}_r}\sum _{s,J} [J]
W_{rsrs}^{JT} (1+\delta _{rs})
-\frac{1}{{\mathcal N}} \sum _{tu,J} [J]
W_{tutu}^{JT} (1+\delta _{tu}),
\label{ispeTS} \\
W_{rstu}^{JT}(2)=W_{rstu}^{JT}-(W_c^T +\frac{\lambda _r^T +\lambda _s^T}
{{\mathcal N}+2(-1)^T})
\delta _{rt}\delta _{su}.
\label{W2TS}
\end{eqnarray}
  In order to calculate the correlation coefficient $\zeta ^{n,T}$ and the
variance $\sigma ^{n,T}$, the following quantities are needed (the 
sum $\sum^*$ goes over $r\leq s$,
$t\leq u$ and $J$),
\begin{eqnarray}
&\langle H^\dagger H^\prime\rangle ^{n,T}-\langle H^\dagger \rangle
^{n,T}\langle
H^\prime \rangle ^{n,T}=
p_1(T)\sum _{r} {\mathcal N}_r \tilde \varepsilon _{r}\tilde
\varepsilon ^\prime _{r}
+\sum _{r,\tau } p_1(n,T,\tau ){\mathcal N}_r
[\tilde \varepsilon _{r}{\lambda^\prime} _{r}^\tau+
\tilde \varepsilon ^\prime _{r}\lambda _{r}^\tau ]
\nonumber \\
&+\sum _{r,\{\tau _1,\tau _2 \}} p_1(n,T,\tau _1,\tau _2){\mathcal N}_r
[\lambda _{r}^{\tau _1}{\lambda ^\prime }_{r}^{\tau _2} +
      \lambda _{r}^{\tau _2} {\lambda ^\prime }_{r} ^{\tau _1}] \nonumber \\
&+\sum _\tau  p_2(n,T,\tau )\frac{2}{{\mathcal N} ({\mathcal N}+(-1)^\tau )}
\textstyle{\sum ^*}
[J] W_{rstu}^{J\tau } (2) {W^\prime }_{rstu}^{J\tau } (2),
\label{<HH'>nT}
\end{eqnarray}
where $\tau $ is $0$ or $1$, and the set $\{\tau _1,\tau _2\}$
is $\{0,0\}, \{0,1\}$ or $\{1,1\}$. The propagator functions
are derived in
\cite{French69,HechtDraayer74} to be,
\begin{align}
p_1(T)=\textstyle{
\frac{n({\mathcal N}+2)({\mathcal N}-\frac{n}{2})-2{\mathcal N}T(T+1)}
{{\mathcal N}({\mathcal N}-1)({\mathcal N}+1)}
}
\end{align}
\begin{align}
p_1(n,T,\tau )&=
\textstyle{
\frac{
4\dimN T(T+1) (1-n) (1-(-1)^\tau )+
(\dimN +2)({\mathcal N}-\frac{n}{2}) [(2\tau 
+1)n(n+2(-1)^\tau)-4T(T+1)(-1)^\tau ]
}{4\dimN (\dimN -1)(\dimN +1)(\dimN +2(-1)^\tau )}
}
\nonumber
\end{align}
\begin{align}
p_1(n,T,\tau _1,\tau _2)=
\textstyle{
\frac{1}{8(\dimN -1)(\dimN +1)(\dimN -2)(\dimN +2(-1)^{\tau _1})}
}
\{4\dimN T(T+1) (n-1) \times \nonumber\\
(\dimN \textstyle{-} 
2n\textstyle{+}4)(1\textstyle{-}(\textstyle{-}1)^{\tau _1})+[(2\tau _1
+1)n(n+2(-1)^{\tau _1})-4T(T+1)(-1)^{\tau _1}]
\times
\nonumber\\ [(2{\tau _2}+1)(n+2(-1)^{\tau _2})(\dimN 
-\frac{n}{2})\half+T(T+1)(-1)^{\tau _2}]
[\dimN-2(-1)^{\tau _2}]
\}
\nonumber
\end{align}
\begin{align}
p_2(n,T,\tau =0)&= \textstyle{\frac{[n(n+2)-4T(T+1)]
[(\dimN -\frac{n}{2})(\dimN -\frac{n}{2}+1)-T(T+1)]}{8\dimN (\dimN-1)}
}
\nonumber
\end{align}
\begin{align}
p_2(n,T,\tau =1)= \textstyle{
\frac{1}{\dimN (\dimN +1)(\dimN -2)(\dimN -3)}
}
\{\half T^2 &(T+1)^2(3\dimN ^2-7\dimN +6) \nonumber\\
+\frac{3}{8}n(n-2)(\dimN -\frac{n}{2})(\dimN -\frac{n}{2}+&1)(\dimN
+1)(\dimN +2)
\nonumber\\
+\half T(T+1)[(5\dimN -3)(\dimN +2)n(\frac{n}{2}-\dimN) +&
\dimN (\dimN -1)(\dimN +1)(\dimN +6)]\}.
\nonumber
\end{align}

For the \Spn{4} interaction, the average two-body interaction is
expressed in terms of the model parameters in the scalar distribution as,
$W_c=-\frac{3G_0}{\binom{\dimN}{2}}+\frac{3E_0}{2(\dimN -1)}-C$, and 
in the isospin-scalar case for
a given isospin value as, $W_c^T=-\frac{G_0}{\binom{\dimN}{2}}\delta 
_{T1} +E_0[(-1)^T+\half ]-C$.
The pure one-body part of the \Spn{4} Hamiltonian is zero for a 
singe-$j$ orbit (by definition) as
well as for the \fpg major shell (by assumption of constant 
interaction strengths). The pure two-body
matrix elements, $W_{rstu}^{JT}(2)$, and hence the correlation
coefficients involving \Hsp, are then independent of the $C$ 
parameter in the scalar case and of
the C and $E_0$ parameters in the isospin-scalar case.

\end{document}